\documentclass{article}
\usepackage{spconf,amsmath,graphicx,hyperref}
\usepackage{xcolor} 
\usepackage{cite}
\usepackage{algorithm}
\usepackage[numbers,sort&compress]{natbib}
\usepackage{comment}
\usepackage{subfigure}
\usepackage{adjustbox}
\usepackage{caption}
\usepackage{subcaption}

\usepackage{booktabs}
\usepackage{multicol, multirow}
\usepackage{color, colortbl}

\usepackage{booktabs} 
\usepackage{siunitx}  
\usepackage{multirow} 
\usepackage{makecell}

\title{T-Mimi: A Transformer-based Mimi Decoder for Real-Time On-Phone TTS}
\name{
\begin{tabular}{@{}c@{}}
 Haibin Wu, Bach Viet Do, Naveen Suda, Julian Chan, Madhavan C R, Gene-Ping Yang \\ 
 \textit{Yi-Chiao Wu, Naoyuki Kanda, Yossef Adi, Xin Lei, Yue Liu, Florian Metze, Yuzong Liu}
\end{tabular}
}
\address{Meta, USA}
\begin{document}
%
\maketitle

\begin{abstract}
Neural audio codecs provide promising acoustic features for speech synthesis, with representative streaming codecs like Mimi providing high-quality acoustic features for real-time Text-to-Speech (TTS) applications.
However, Mimi's decoder, which employs a hybrid transformer and convolution architecture, introduces significant latency bottlenecks on edge devices due to the the compute intensive nature of deconvolution layers which are not friendly for mobile-CPUs, such as the most representative framework XNNPACK~\cite{google_xnnpack}. 
This paper introduces T-Mimi, a novel modification of the Mimi codec decoder that replaces its convolutional components with a purely transformer-based decoder, inspired by the TS3-Codec architecture. 
This change dramatically reduces on-device TTS latency from 42.1ms to just 4.4ms. 
Furthermore, we conduct quantization aware training and derive a crucial finding: the final two transformer layers and the concluding linear layers of the decoder, which are close to the waveform, are highly sensitive to quantization and must be preserved at full precision to maintain audio quality. 
\end{abstract}


\section{Introduction}
\label{sec:intro}

Recent advances in neural audio codecs~\cite{guo2025recent,arora2025landscape,wu2024towards,mousavi2025discrete} have fundamentally transformed the intermediate representations used in modern text-to-speech (TTS) systems. 
While mel-spectrograms have long served as the de facto standard acoustic feature, representations derived from models such as EnCodec \cite{defossez2022high}, DAC \cite{kumar2023high} and Mimi~\cite{defossez2024moshi} have emerged as a superior alternative. These modern codec features, whether in the form of discrete tokens or continuous vectors, offer exceptional compression ratios while preserving perceptually vital information, sometimes leading to higher audio quality than traditional mel-spectrogram-based approaches~\cite{mousavi2025discrete}. 
This paradigm shift effectively redefines the TTS pipeline: the new synthesis model now predicts codec features instead of mel-spectrograms, and the neural codec's own high-fidelity decoder replaces the traditional vocoder, serving as the final stage of generating waveforms from acoustic features.

Among the many NACs, the Mimi codec stands out as an excellent choice for building streaming TTS systems. 
Mimi offers several key advantages: First, it skillfully disentangles the audio signal into separate ``semantic" and ``acoustic" feature streams. 
This separation allows a TTS model to exert fine-grained control over the synthesized content (what is said) and the speech style (how it is said). 
Second, its decoder can reconstruct high-quality, natural-sounding audio from these tokens under a really low frame rate (12.5HZ). 
Most importantly, Mimi supports streaming, enabling real-time encoding and decoding of audio segments. 
The streaming decoder is crucial for streaming TTS applications that require immediate responses.

However, a significant challenge emerges when deploying advanced models like Mimi onto mobile or edge devices. 
Mimi's decoder relies on de-convolution (or transposed convolution) layers for upsampling to reconstruct the final audio waveform. 
While convolutions are highly parameter-efficient due to weight sharing, they can still be computationally intensive \footnote{With the same parameter amount, the transformer codec requires 1/7 computation compared with CNN based codec \cite{wu2024ts3}.}.
While these layers can be optimized efficient in theory, they are often computationally inefficient in mobile inference frameworks like XNNPACK \footnote{https://github.com/google/XNNPACK}, leading to significant latency. 
These frameworks often have better optimization for transformers than de-convolution layers.
This latency issue becomes the primary bottleneck for achieving a smooth, real-time TTS experience on phones.

TS3-Codec~\cite{wu2024ts3}, being purely Transformer-based and convolution-free, establishes a new benchmark for efficiency. 
It not only greatly outperforms CNN-based NACs when matched for parameter size, but also delivers comparable performance at a fraction of the computational cost—requiring a mere 13\% of the resources.
Inspired by TS3-Codec, this paper introduces T-Mimi, a modified version of the Mimi codec featuring a Transformer-only decoder architecture. 
By replacing the slow de-convolution layers with a mobile-friendly architecture composed of Transformers and linear layers, we directly address the on-device latency bottleneck.
We also applied quantization aware training (QAT) to further reduce the model's storage footprint and computational requirements, making it better suited for resource-constrained devices.
Our primary contributions are as follows:
(1). We designed T-Mimi, drastically reducing on-device TTS decoding latency from 42.1ms to just 4.4ms, enabling true real-time TTS performance. A latency of 42.1 ms to generate a single 80ms audio frame is a significant bottleneck to achieve real-time processing. When factoring in the processing time required for other essential TTS components, such as the front-end and acoustic models, along with potential resource contention on the device, this delay makes achieving a seamless, real-time audio experience highly improbable.
(2). Through quantization aware training, we discovered a crucial rule: the closer a layer is to the final waveform, the worse the audio quality becomes after it is quantized. Consequently, we found that the final layers of the decoder, specifically the last two Transformer layers and the two linear layers important for audio generation. They should be preserved in full precision to maintain audio fidelity.

We take Mimi codec as a representative case study. Our methodology is not limited to Mimi codec, and provides a generalizable framework for optimizing other convolution-based neural audio codecs for efficient on-device deployment.

\section{Background}
\label{sec:realted}

\subsection{Codec feature as a good medium for TTS}

A variety of state-of-the-art TTS models demonstrate the promise of neural audio codec features as acoustic representations. 
This is exemplified by a variety of state-of-the-art models, such as VALLE~\cite{wang2023neural}, VoiceStar~\cite{peng2025voicestar}, which leverage Encodec features, AudioLM~\cite{borsos2023audiolm} and SpearTTS~\cite{kharitonov2023speak}, which utilize SoundStream tokens~\cite{zeghidour2021soundstream}, and Spark-TTS~\cite{wang2025spark}, builts upon BiCodec features. 
To generate these acoustic features from text or semantic tokens, these systems employ a range of acoustic modeling techniques, including diffusion models~\cite{wang2024maskgct}, Tacotron-based architectures~\cite{mousavi2025discrete}, and depth-transformer-based methods~\cite{yang2023uniaudio}.
Ultimately, regardless of the specific acoustic models and codec features, every system converges on the codec's decoder as the final, essential step to transform the predicted codec features into the output waveform.
Our focus in this paper is to develop an on-phone streaming and low-latency codec decoder.

\subsection{On-phone TTS requirements}
On-phone streaming TTS systems demand a delicate balance of competing requirements: low latency on constrained hardware, a streaming-capable architecture, and high-fidelity audio reconstruction. Additionally, a minimal storage footprint is essential, making quantization a necessity. To this end, quantization aware training is employed to preserve audio quality during model compression. The Mimi codec is a compelling foundation as it already provides a streaming architecture and high-quality reconstruction, making it an ideal candidate for further on-device optimization.

\subsection{Quantization aware training}
Quantization is an important technique for optimizing the efficiency of deep learning models, enabling significant reductions in both memory footprint and inference latency. 
To mitigate the performance degradation typically associated with post-training quantization, we employ quantization aware training, which simulates the effects of low-precision arithmetic during the training process.

\section{Method}
\label{sec:method}
\begin{figure}[ht]
    \centering
    \includegraphics[width=0.8\columnwidth]{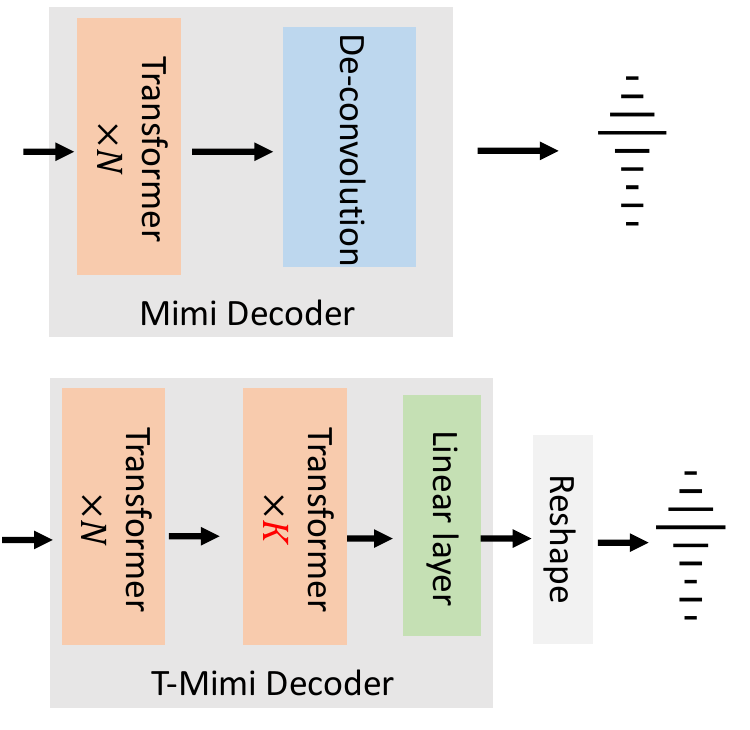}
    \caption{Model architectures for the original Mimi decoder (upper) and the T-Mimi decoder (bottom).}
    \label{fig:t_mimi}
    \vspace{-10pt}
\end{figure}
\subsection{Architecture}
The model architectures of the original Mimi decoder and the proposed T-Mimi decoder are shown in the upper and bottom of Figure~\ref{fig:t_mimi}.
The standard Mimi decoder utilizes a hybrid design of eight Transformer layers and subsequent de-convolution layers for upsampling. Inspired by our TS3-Codec, we developed a Transformer-only alternative by replacing the de-convolution module with four additional Transformer layers (with fixed window streaming self-attention) and two linear layers, while keeping the total parameter count constant. These linear layers, with the first having a bias and the second without, perform the upsampling, and the resulting waveform segments are concatenated directly, omitting any overlap-and-add. 

From our intial experiments, we found that a deeper, 12-layer architecture had better performance than a wider 8-layer one. One possible reason is that the 12-layer setting allows us to leverage the original eight pre-trained Mimi Transformer layers, providing a powerful initialization.

\subsection{Training}

During training, the T-Mimi encoder is fixed, and our T-Mimi decoder model is trained using a composite loss of GAN losses and reconstruction losses. 
The final objective is a weighted sum of four key components. 
First, a multi-scale mel-spectrogram reconstruction loss (L1 distance) serves as the primary reconstruction loss, with a weight of 2.0. Second, a least-squares GAN loss and a feature matching loss are used to improve the audio quality, each with a weight of 4.0. 
These rely on a Multi-Scale STFT Discriminator architecture~\cite{kumar2023high}. Finally, an L1 loss term with a weight of 0.1 is applied. 

We employ a two-stage training strategy to improve the audio quality. 
In the initial stage, the model is trained with the full composite loss until a satisfactory level of convergence is achieved. Following this, we initiate a fine-tuning phase where the T-Mimi decoder is trained exclusively with the feature matching loss. 
The continual training enhances the subjective perceptual audio quality.

Additionally, we observed that the model tended to generate low-level noise in silent regions of the audio. To mitigate this artifact, we introduced a data augmentation technique. For 10\% of the training samples, we prepended and appended segments of pure silence to the audio clips. This simple yet effective trick forces the model to explicitly learn a robust representation of silence and proved highly successful in suppressing noise artifacts during silent intervals.

\subsection{Quatization aware training setup}

For the T-Mimi decoder, we utilized the TorchAO library \footnote{https://github.com/pytorch/ao} to implement and evaluate various quantization schemes, including 4-bit group-wise and 8-bit per-channel quantization for weights, alongside 8-bit dynamic quantization for activations. 
We hypothesized that the final layers of the decoder would be particularly sensitive to precision loss, given the high fidelity required to reconstruct the 24kHz audio waveform.
Through extensive ablation studies (Table~\ref{tab:qat}), our hypothesis was confirmed. 
We determined that the optimal trade-off between model quality, size, and latency was achieved by preserving the final two layers (two Transformer and two linear layers) in their original full precision (FP32) while quantizing the rest of the network. 
This selective quantization strategy was then incorporated into a final QAT phase to further refine the model's performance.

\section{Experiments}
\label{sec:experiment}
\subsection{Experimental setup}
\subsubsection{Training setup}
For training, we use an in-house speech dataset of 5 million hours. To get a fair baseline, we fine-tune the open-source Mimi model on the same data (the model is denoted as Mimi-FT-32-bit), and train our proposed T-Mimi-32-bit on the same corpus, both at full 32-bit precision. 
For evaluation, performance is measured on an audio reconstruction task using 100 randomly selected speech samples.

Through a series of ablation studies, we determine the optimal architecture for the T-Mimi decoder. The best-performing configuration employs 12 Transformer layers with a hidden dimension of 2048. We will present the detailed experiments that led to this design choice in the ablation study.
The model is optimized using the Adam optimizer with a learning rate of $5 \times 10^{-4}$. 
Then we apply quantization aware training for the T-mimi.
For quantization aware training, we apply 8-bit per-channel quantization. 
During QAT, the Adam is used with a learning rate of $1 \times 10^{-5}$.

\subsubsection{Evaluation setup}
Our evaluation is based on a set of objective metrics. For audio quality, we measure Short-Time Objective Intelligibility (STOI), wide-band Perceptual Evaluation of Speech Quality (PESQ), and Scale-Invariant Signal-to-Distortion Ratio (SI-SDR) to assess intelligibility, perceptual quality, and reconstruction fidelity, respectively. To evaluate model efficiency, we also report the number of parameters (M), the on-disk model size (MB), and the inference latency (ms).

\subsection{Experimental results}

\subsubsection{CMOS between non-quantized Mimi and T-Mimi}
\begin{table}[ht]
    \centering
    \footnotesize
    \caption{Human CMOS evaluation results between T-Mimi decoder and baseline Mimi decoder} 
    \renewcommand{\arraystretch}{1.2}
    \begin{tabular}{ccc}
      \toprule
      \multirow{2}{*}{\textbf{Model comparison}} & \textbf{Average} & \textbf{ 95\% Confidence} \\
      & \textbf{winrate} & \textbf{Interval} \\
      \midrule
      T-Mimi-32-bit vs. Mimi-FT-32-bit  & $+2.32\%$ & $(-0.70\%, 5.34 \%) $  \\
      \bottomrule
    \end{tabular}

    \label{tab:cmos} 
\end{table}
Human evaluation is the golden standard to evaluate the audio quality, thus we compare the Mimi-FT-32-bit and T-Mimi-32-bit by human evaluation.
Table \ref{tab:cmos} shows the CMOS (Comparative Mean Opinion Score) human evaluation results. 
CMOS is a commonly used metric in the speech industry for evaluating the relative voice quality of two systems (Mimi-FT-32-bit, and T-Mimi-32-bit). 
In our evaluation, 200 pairs of audio samples were assessed, with each pair rated by 10 qualified raters. 
The raters compared each pair and indicated which sample sounded better. 
The reported CMOS gap reflects the average difference in opinion scores between the two systems. 
The result shows no significant difference between Mimi-FT-32-bit, and T-Mimi-32-bit, which means these two systems are on-par.
And in the following experiments, we conduct quantization aware training of the T-Mimi-32-bit in Table~\ref{tab:cmos}.
\begin{table}[ht]
    \centering
    \footnotesize
    \caption{Comparison of QAT settings, storage (in MB), and audio quality metrics. Run for 50k steps to select models.}
    \renewcommand{\arraystretch}{1.1} 
    \rowcolors{3}{gray!10}{white} 
    \begin{tabular}{
        >{\raggedright\arraybackslash}p{3.5cm} 
        >{\centering\arraybackslash}p{0.8cm}      
        >{\centering\arraybackslash}p{0.8cm}    
        >{\centering\arraybackslash}p{0.8cm}    
        >{\centering\arraybackslash}p{0.8cm}    
    }
        \toprule
        \textbf{QAT Setting} & \textbf{Storage} & \textbf{PESQ} & \textbf{STOI} & \textbf{SISDR} \\
        \midrule
        T$_{\text{1--12, 4bit}}$ -- L$_{\text{4bit}}$ & 20.4 & 2.32 & 0.96 & 15.82 \\
        T$_{\text{1--12, 8bit}}$ -- L$_{\text{8bit}}$ & 40.8 & 2.74 & 0.97 & 18.42 \\
        T$_{\text{1--12, 8bit}}$ -- L$_{\text{32bit}}$ & 50.3 & 2.81 & 0.98 & 18.30 \\
        T$_{\text{1--11, 8bit}}$ -- T$_{\text{12, 32bit}}$ -- L$_{\text{32bit}}$ & 59.2 & 2.96 & 0.98 & 19.87 \\
        T$_{\text{1--10, 8bit}}$ -- T$_{\text{11--12, 32bit}}$ -- L$_{\text{32bit}}$ & 68.7 & 2.99 & 0.98 & 19.62 \\
        T$_{\text{1--9, 8bit}}$ -- T$_{\text{10--12, 32bit}}$ -- L$_{\text{32bit}}$ & 78.2 & 3.04 & 0.98 & 20.10 \\
        \bottomrule
    \end{tabular}
    \label{tab:qat}
\end{table}

\subsubsection{Quantization aware training}
To further reduce model storage and latency while preserving the 32-bit performance, we investigate the efficacy when applying QAT following full-precision model training.
The encoder is frozen during training.
Table \ref{tab:qat} summarizes six different QAT strategies that assign mixed precision to the model layers.
Note, in Table~\ref{tab:qat}, all models are only trained for 50k steps to save computation and select models.
In these experiments, the initial layers are set to a lower precision (4-bit or 8-bit), while the precision of the final layers is varied.
While the 4-bit model exhibits a decline in audio quality, the 8-bit model maintains strong despite achieving a 75\% reduction in storage compared to the 32-bit model.
The perceptual quality of the 8-bit model can be further improved by only increasing the precision of the final layers.
To maintain both good audio quality and small storage, we choose model T$_{\text{1--10, 8bit}}$ -- T$_{\text{11--12, 32bit}}$ for continue training.
The final QAT model after full training steps (in Table~\ref{tab:latency_benchmarking}) achieves a PESQ of 3.16, which is very close to the non-quantized model's a PESQ of 3.21, and reducing the storage from 163.2MB to 68.7MB.

\begin{table}[ht]
    \centering
    \footnotesize
    \caption{Average latency and storage requirements for generating 80-ms audio segments. CNN-Mimi (win=5) indicates the baseline model.}
    \renewcommand{\arraystretch}{1.2}
    \label{tab:latency_benchmarking}
    \begin{tabular}{@{}lccc@{}}
        \toprule
        & \textbf{CNN-Mimi (win = 5)} & \textbf{CNN-Mimi (win = 2)} & \textbf{T-Mimi} \\
        \midrule
        \textbf{Latency} & 42.1 ms & 18.0 ms & 4.4 ms \\
        \textbf{Storage} & 81.0 MB & 81.0 MB & 68.7 MB \\
        \bottomrule
    \end{tabular}
\end{table}

\subsubsection{On-phone latency benchmarking results}
We compare the inference latency of the proposed T-Mimi decoder against the deconvolution-based baseline. 
The baseline is configured to use a context-window of 5 for the CNN layers. We also show the numbers of reducing that window size to 2. While reducing the context-window size reduces the latency, it can degrad the generated audio quality. The benchmarking is performed on a Samsung Galaxy S22 smartphone, with the Mimi decoder running as part of a TTS system. We report the average latency of generating 80-ms audio chunks.
\begin{table}[]
    \centering
    \footnotesize
    \caption{Ablations for layer selection and linear dimension}
    \renewcommand{\arraystretch}{1.2} 
    \rowcolors{3}{white}{gray!10} 
    \begin{tabular}{ccccccc}
        \toprule
        \multirow{2}{*}{\textbf{Layer}} & \textbf{Linear} & \textbf{Storage} & \textbf{Para \#} & \multirow{2}{*}{\textbf{PESQ}} & \multirow{2}{*}{\textbf{STOI}} & \multirow{2}{*}{\textbf{SISDR}} \\
        & \textbf{dim} & \textbf{(MB)} & \textbf{(M)} & & & \\
        \midrule
        8  & 2048 & 131.4 & 28.2 & 2.61 & 0.96 & 16.10 \\
        12 & 2048 & 163.2 & 40.8 & 2.95 & 0.98 & 19.37 \\
        12 & 3072 & 169.2 & 42.3 & 2.96 & 0.98 & 19.41 \\
        16 & 2048 & 207.0   & 53.4 & 3.07 & 0.98 & 19.91 \\
        \bottomrule
    \end{tabular}
    \label{tab:layer_selection}
\end{table}
As shown in Table~\ref{tab:latency_benchmarking}, the proposed T-Mimi decoder requires only 4.4 ms for each 80ms audio chunk, showing a 9.6x speed-up over the baseline, making true real-time TTS experience. 
Notably, the CNN-Mimi model was evaluated without undergoing quantization aware training owing to implementation constraints within the CNN QAT libraries. Applying quantization to a model not specifically optimized for it is known to cause a loss in performance, a vulnerability that is particularly acute for generative models.

\subsubsection{Ablation studies}

In Table \ref{tab:layer_selection}, we present the trade-off between model storage and performance by varying both the model layers and the dimension of the final linear layers.
These models are trained for only 90k steps for model selection.
First, expanding the model from 8 to 12 layers lead to substantial improvements across all evaluation metrics, demonstrating that 8-layer models might be too shallow to effectively capture subtle audio features.
For the 12-layer model, increasing the linear dimension to 3072 results in modest improvement, with a storage overhead of just over 6 MB.
On the other hand, increasing the number of layers to 16 yields also additional gains, though the improvements are less pronounced compared to the jump from 8 to 12 layers.
Taking these trade-offs into consideration, we therefore select the configuration with 12 layers and a linear dimension of 2048 for all experiments throughout the paper as it demonstrate good performance considering resource constraints.

\section{Conclusion}
This paper introduces T-Mimi for on-phone TTS senario, a purely transformer-based decoder that serves as a highly efficient alternative to the conventional hybrid transformer-convolution design of the Mimi codec. 
By replacing the computationally intensive de-convolution layers, T-Mimi achieves a 9.6x reduction in on-device inference latency, bringing it down to just 4.4ms and enabling genuine real-time TTS performance on mobile devices.
Furthermore, the study on quantization aware training reveals a critical principle for model compression: layers closest to the final waveform output, are exceptionally sensitive to reduced precision. 
To preserve high-fidelity audio, these final layers must be maintained at full precision.

\bibliographystyle{IEEEbib}
\bibliography{strings,refs}

\end{document}